\documentclass[12pt]{iopart}
\usepackage{graphicx}
\usepackage{cite}
\usepackage[colorlinks,linkcolor=red,anchorcolor=green,citecolor=blue,CJKbookmarks=True]{hyperref}

\begin{document}

\title[Author guidelines for IOP Publishing journals in \LaTeXe]{Photon-pair blockade in a Josephson-photonics circuit with two nondegenerate microwave resonators}

\author{Sheng-li Ma$^1$,Ji-kun Xie$^1$,Ya-long Ren$^1$, Xin-ke Li$^{2,*}$ and Fu-li Li$^{1,*}$}

\address{$^1$ MOE Key Laboratory for Non-equilibrium Synthesis and Modulation of
Condensed Matter, Shaanxi Province Key Laboratory of Quantum Information and
Quantum Optoelectronic Devices, School of Physics, Xi’an Jiaotong
University, Xi’an 710049, China}
\address{$^2$ School of Mathematics, Physics and Optoelectronic Engineering,
Hubei University of Automotive Technology, Shiyan 442002, China}
\ead{20210064@huat.edu.cn and flli@mail.xjtu.edu.cn}

\vspace{10pt}

\begin{abstract}
We propose to generate photon-pair blockade in a Josephson-photonics circuit
that consists of a dc voltage-biased Josephson junction in series with a
superconducting charge qubit and two nondegenerate microwave resonators. The
two-level charge qubit is utilized to create anticorrelations of the charge
transport; that is, the simultaneous Cooper pair tunneling events are
inhibited. When the Josephson frequency matches the sum of resonance
frequencies of the charge qubit and the two resonators, we demonstrate that
the two resonators can release their energies in the form of antibunched
pairs of two strongly correlated photons. The present work provides a
practical way for producing a bright microwave source of antibunched photon pairs,
with potential applications ranging from spectroscopy and metrology to
quantum information processing.
\end{abstract}

\begin{indented}
\item[]\today
\end{indented}

%
%
%
%
%

\section{Introduction}

Solid-state superconducting circuit studies the interaction between
artificial atoms and quantized electromagnetic fields in the microwave
frequency domain \cite{RevModPhys93/025005}. This architecture has emerged as one of the leading platforms for realizing quantum computation and simulation \cite{Arute2019,PhysRevLett127/180501}. Commonly,
superconducting qubits are used for encoding quantum information, with
superconducting resonators acting as date buses. However, the efficient
generation, manipulation, and transmission of nontrivial quantum states in a
linear resonator are also crucial to different kinds of quantum information
tasks \cite{Gu2017}. The quantum states of a harmonic oscillator are extraordinarily rich,
but are hard to access due to the infinitely equally spaced energy levels.
This difficulty can be overcome by interposing a nonlinear artificial atom,
and various quantum states in a resonator can be synthesized via the
deliberate use of classical control signals, such as Fock state \cite{Hofheinz2008} and Schr\"{o}dinger cat state \cite{Ofek2016}.

In recent years, the Josephson-photonics circuit of a dc voltage-biased
Josephson junction in series with a microwave resonator has emerged as an
alternative tool for efficient on-chip generation of coherent microwave
photons \cite{M1,M2,M3,M4,M5,M6,M7,M8,M9,M10}. It relies on the exceptionally strong nonlinearity of light-charge interaction, and eliminates the need for any
microwave drives\cite{D1,D2,D3,D4,Kubala2020}. The Josephson
junction acts as a highly nonlinear driving element; that is, the inelastic
Cooper pair tunneling across a junction can convert different numbers of
photons from an easily controlled bias voltage source into a microwave
cavity \cite{PhysRevLett247001,PhysRevLett247002}. By modulating the charge tunneling effect via a specifically tailored electromagnetic environment, the nonclassical microwave light can be generated \cite{non1,non2,non3,non4,non5,Dambach2019}. The resulting quantum electrodynamics of this simple circuit
has been demonstrated to realize Josephson junction lasers \cite{L1,Cassidy2017,L2}, single-photon sources \cite{S0,S1,S2,S3}, multi-photon sources \cite{mul1,mul2,mul3}, and near quantum-limited amplifiers \cite{Jebari2018}.

In parallel, a significant development in this field is to connect Josephson
junctions with multiple resonators for various quantum technological
applications, such as the implementations of entangled quantum microwaves \cite{Dambach2017}
and microwave single-photon detectors \cite{SD}. While the simplest case is one voltage-biased junction in series with two cavities of incommensurate
frequencies. When the bias voltage is tuned to match the energy required to
simultaneously produce one photon in each cavity for a single Cooper pair
traversing the circuit, the system can be effectively reduced as a
nondegenerate parametric amplifier \cite{F1,F2,F3}. It allows the continuous emission of
correlated photon pairs. Recent experiments have observed the amplitude
squeezing \cite{squeeze} and entanglement \cite{entangle} of these output microwave beams. Especially, when the cavities possess the impedance of 4.1 k$\Omega$, there is no matrix element for a transition between the one and two photon states \cite{PhysRevLett247002,S2}. So, the two cavities can be regarded as two-level systems, leading to an antibunched photon-pair source \cite{F1}. However, the fabrication of coplanar waveguide resonators with such high
impedances is highly challenging as the standard cavity designs only yields
characteristic impedances of the order of 100 $\Omega$.

To go beyond this limitation, we propose a more practical way to realize
photon-pair blockade by regulating the anticorrelated behavior of the charge
transport via a two-level charge qubit. In our scheme, we study the
Josephson-photonics device of a voltage-biased Josephson junction in series
with a charge qubit and two nondegenerate microwave resonators. The
nonlinear qubit-resonator coupling can be sculpted via the phase difference
across the junction. For each tunneling Cooper pair, the suitably set bias
voltage enables the excitation of charge qubit and the creation of one
photon in each resonator concurrently. Since the charge qubit is an ideal
anharmonic element with two quantum energy levels, the anticorrelations of
the tunneling Cooper pairs can be created, preventing the simultaneous
tunnel events. Combined with the dissipation, we show that the two
resonators can release their energies in the form of antibunched photon
pairs in a controllable manner. Compared with the previous work \cite{F1}, the present
one constitutes a significant step forward; that is, the photon-pair source
can be achieved with the standard coplanar waveguide resonator designs,
eliminating the need for ultrahigh cavity impedance that is not accessible
in current experiment. Our work offers an appealing method for generating a
bright nonclassical source of antibunched pairs of two strongly correlated
photons, required at the core of quantum computing and quantum communication
protocols \cite{P0,P1,P2,P3}.

\begin{figure}[tbh]
\centering \includegraphics[width=\linewidth]{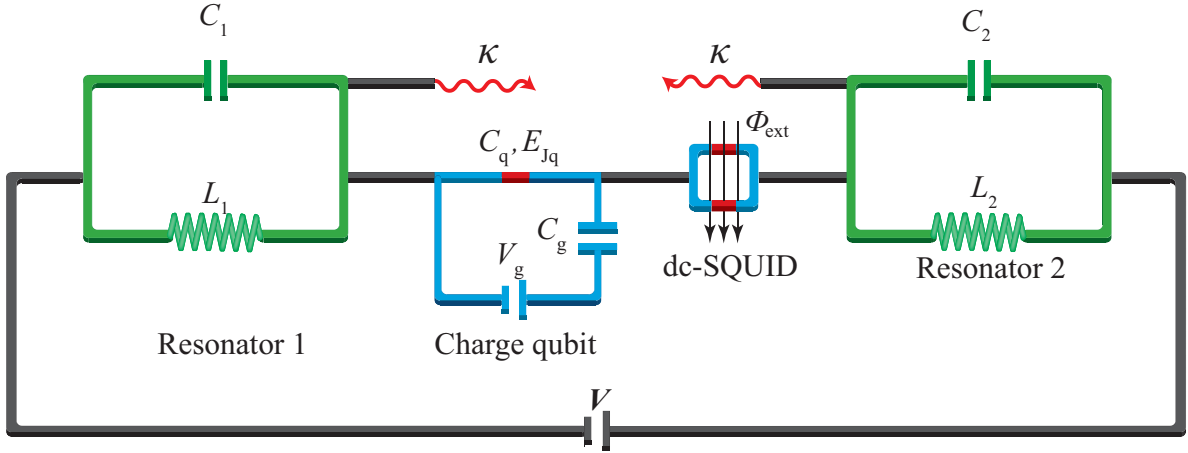}
\caption{Schematic diagram of the proposed experimental setup that consists
of a dc-SQUID coupled to a charge qubit and two nondegenerate $LC$
resonators. A dc bias voltage $V$ is applied across all the elements, and a
flow of Cooper pairs across the circuit can pump both the qubit and the two $%
LC$ resonators. In addition, each resonator is coupled to a transmission line with the coupling strength $\kappa$, which is used for collecting the emitted photons.}
\label{fig.1}
\end{figure}

\section{Model}

As shown in Figure 1, we investigate the Josephson-photonics circuit of a
voltage-biased dc superconducting quantum interference device (dc-SQUID) in
series with a charge qubit and two nondegenerate $LC$ resonators. We focus on the
situation that the bias voltage $V$ is smaller than the gap voltage, and no
quasi-particle excitation can be produced in the superconducting electrodes.
So, the quantum transport of Cooper pairs through the circuit will supply
energies to both the charge qubit and the two $LC$ resonators. The model
Hamiltonian describing the entire setup takes the form (see Appendix)

\begin{eqnarray}\label{1}
H_{T} &=&E_{c}(n_{q}-n_{g})^{2}-E_{Jq}\cos \phi _{q}+\sum_{j=1,2}[\frac{%
q_{j}^{2}}{2C_{j}}+(\frac{\hbar }{2e})^{2}\frac{\phi _{j}^{2}}{2L_{j}}]
\nonumber \\
&&\quad -E_{J}\cos \phi _{J}-2en_{J}(V-V_{q}-V_{R1}-V_{R2}).
\end{eqnarray}
The first two terms denote the part of charge qubit, the third term describes the part of two $LC$ resonators, and the last two terms represent the part of dc-SQUID, where $V_{q}=-\hbar\dot{\phi}_{q}/2e$ is the voltage drop at the qubit, and $V_{Rj}=-\hbar\dot{\phi}_{j}/2e$ is the voltage drop at the $j$-th resonator.

To exclude the Cooper pair number $n_{J}$, we perform a unitary
transformation $U(t)=\exp [i(\omega _{J}t+\phi _{q}+\phi _{1}+\phi
_{2})n_{J}]$ on the full Hamiltonian, where $\omega _{J}=2eV/\hbar $ is the
Josephson frequency. Then, we can obtain

\begin{eqnarray}
\tilde{H}_{T} &=&U^{^{\dag }}(t)H_{T}U(t)+i\hbar \frac{dU^{^{\dag }}(t)}{dt}%
U(t)  \nonumber \\
&=&E_{c}(\tilde{n}_{q}-n_{g})^{2}-E_{Jq}\cos \eta _{q}+\sum_{j=1,2}[\frac{%
\tilde{q}_{j}^{2}}{2C_{j}}+(\frac{\hbar }{2e})^{2}\frac{\phi _{j}^{2}}{2L_{j}%
}]  \nonumber \\
&&\quad -E_{J}\cos (\omega _{J}t+\phi _{q}+\phi _{1}+\phi _{2}),
\end{eqnarray}%
where $\tilde{n}_{q}=n_{q}+n_{J}$, $\tilde{q}_{j}=q_{j}+2en_{J}$ are the
transformed number and charge operators, arising from the charge
fluctuations regard to the flow of Cooper pairs through the dc-SQUID. As
described by the last term in $\tilde{H}_{T},$ the nonlinear coupling
between the charge qubit and the two cavities is established via the phase
difference across the junctions of dc-SQUID.

When the charge qubit is operated at the degeneracy point with $n_{g}=1/2$, we can quantize the excitations in the two resonators and
qubit, and the Hamiltonian of the whole circuit will yield (hereafter let $%
\hbar =1$)
\begin{equation}
H=\frac{1}{2}\delta \sigma _{z}+\sum_{j=1,2}\omega _{j}a_{j}^{\dag
}a_{j}-E_{J}\cos [\omega _{J}t+\phi _{q}+\sum_{j=1,2}2\lambda
_{j}(a_{j}^{\dag }+a_{j})].
\end{equation}%
Note that the charge qubit has been reduced as a two-level system containing
an excited state $|e\rangle $ and a ground state $|g\rangle $ under the
basis of charge states \cite{Bouchiat1998,Makhlin1999,Nakamura1999,Jos}, i.e., $\sigma _{z}=|e\rangle \langle e|-|g\rangle \langle
g|$ is the Pauli matrix, and $\delta =E_{Jq}$ is the energy splitting. $a_{j}^{\dag
}$ ($a_{j}$) is the photon creation (annihilation) operator of the $j$th
resonator ($[a_{j},a_{j}^{\dag }]=1$), and $\omega _{j}=1/\sqrt{L_{j}C_{j}}$
is the corresponding resonance frequency. The parameter $\lambda _{j}=\sqrt{%
\pi Z_{j}/R_{K}}$ quantifies the amplitude of the cavity's zero-point
displacement, and characterizes the coupling between the oscillator and the
tunnel junction ($Z_{j}=\sqrt{L_{j}/C_{j}}$ is the characteristic impedance,
and $R_{K}=h/e^{2}$ is the resistance quantum).

\section{Photon-pair blockade}

\subsection{Anticorrelations of the charge transport}

In this section, we will illustrate the procedure for the realization of
photon-pair blockade in the aforementioned two-mode superconducting circuit.
The central idea is to create anticorrelations of the charge transport, which gives rise to the desired antibunching of
the photon pairs leaking out of the two microwave resonators.

To this end, we start to derive the effective Hamiltonian of the system,
which helps to uncover the mechanism of our scheme. In the interaction
picture with respect to the frame rotating $\exp (-iH_{0}t)$, where $H_{0}=%
\frac{1}{2}\delta \sigma _{z}+\omega _{1}a_{1}^{\dag }a_{1}+\omega
_{2}a_{2}^{\dag }a_{2}$ is the free Hamiltonian, we can obtain
\begin{equation}
H_{I}=-\frac{E_{J}}{4}e^{i\omega _{J}t}\left( \sigma _{+}e^{i\delta
t}-\sigma _{-}e^{-i\delta t}-\sigma _{z}\right) D\left[ \alpha _{1}(t)\right]
\times D\left[ \alpha _{2}(t)\right] +h.c.,
\end{equation}%
where we have exploited the formula $e^{i\phi _{q}}=(\sigma _{+}-\sigma
_{-}-\sigma _{z})/2$, and $\sigma _{+}=|e\rangle \langle g|$, $\sigma
_{-}=|g\rangle \langle e|$ are the spin-ladder operators. In addition, $D%
\left[ \alpha _{j}(t)\right] $ is the cavity displacement operator with the
time-dependent amplitude $\alpha _{j}(t)$, which is defined in terms of the
photon creation and annihilation operators as
\begin{equation}
D\left[ \alpha _{j}(t)\right] =e^{[\alpha _{j}(t)a_{j}^{\dag }-\alpha
_{j}^{\ast }(t)a_{j}]},\alpha _{j}(t)=2i\lambda _{j}e^{i\omega _{j}t}.
\end{equation}%
It is now clearly seen that the tunneling of Cooper pairs through a
voltage-biased Josephson junction can not only displace the cavity fields,
but also flip the qubit state. To formulate the desired coupling, we express
$D\left[ \alpha _{j}(t)\right] $ directly in the Fock-state basis \cite{basis}
\begin{equation}
D\left[ \alpha _{j}(t)\right] =\sum_{n=0}^{\infty
}(\sum\limits_{l=0}^{\infty }\beta _{n}^{n+l}(\lambda _{j})|n+l\rangle
\langle n|e^{il\omega _{j}t}+\sum\limits_{l^{\prime }=1}^{\infty }\beta
_{n}^{n+l^{\prime }}(\lambda _{j})|n\rangle \langle n+l^{\prime
}|e^{-il^{\prime }\omega _{j}t})
\end{equation}%
with
\begin{equation}
\beta _{n}^{n+l}(\lambda _{j})=\sqrt{\frac{n!}{(n+l)!}}\left( 2i\lambda
_{j}\right) ^{l}e^{-2\lambda ^{2}}L_{n}^{(l)}(4\lambda _{j}^{2}).
\end{equation}%

Here $\beta _{n}^{n+l}(\lambda _{j})$ is a generalized Frank-Condon factor
that describes an $l$-photon transition rate, and $L_{n}^{(l)}(4\lambda
_{j}^{2})$ is a Laguerre polynomial.

To go a further step, we should tune the bias voltage $V$ to meet the
resonance condition $\omega _{J}=\delta +\omega _{1}+\omega _{2}$, i.e., the
energy provided by the voltage source upon the transfer of a Cooper pair
matches the sum of the excitation energy of qubit and the photon energies of
the two oscillators. In this case, we can retain the resonant terms, but
discard those fast oscillating terms under the rotating-wave approximation
provided that the condition $\delta ,\omega _{j},|\omega _{1}-\omega
_{2}|\gg \frac{E_{J}}{4}|\beta _{n}^{n+l}(\lambda _{1})\beta
_{m}^{m+l^{\prime }}(\lambda _{2})|$ is satisfied. Thus, with all the other
possibilities strongly suppressed, the effective Hamiltonian is derived as
\begin{equation}
H_{eff}=\sum\limits_{n,m=0}^{\infty }H_{nm}
\end{equation}%
with
\begin{equation}
H_{n,m}=g_{eff}^{n,m}|n,m,g\rangle \langle n+1,m+1,e|+h.c.,
\end{equation}%
where $g_{eff}^{n,m}=\frac{E_{J}}{4}\beta _{n}^{n+1}(\lambda _{1})\beta
_{m}^{m+1}(\lambda _{2})$ is the effective coupling strength, and $%
|n,m,g\rangle $ is the tensor product of $|n\rangle \otimes |m\rangle
\otimes |g\rangle $.

\begin{figure}[th]
\centering \includegraphics[width=9cm]{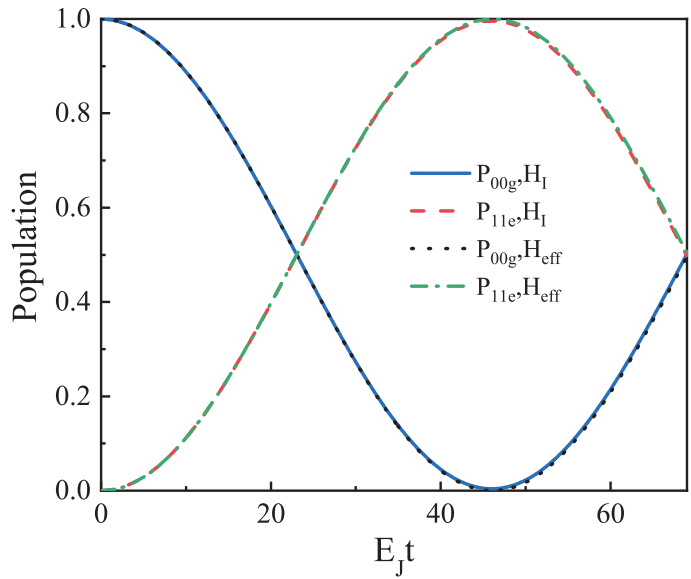}
\caption{(Color online) Time evolution of the state populations $P_{00g}$
and $P_{11e}$. The blue solid and red dashed curves are simulated with the
original Hamiltonian $H_{\mathrm{I}}$, while the black dotted and green
dash-dotted curves are achieved with the effective Hamiltonian $H_{\mathrm{%
eff}}$. The relevant parameters are chosen as $\protect\omega _{1}/2\protect%
\pi =9$ GHz, $\protect\omega _{2}/2\protect\pi =7$ GHz, $\protect\delta /2%
\protect\pi =5$ GHz, $E_{J}/2\protect\pi =0.5$ GHz, and $\protect\lambda %
_{1}=\protect\lambda _{2}=0.2$. }
\label{fig.2}
\end{figure}

The effective Hamiltonian $H_{eff}$ elucidates the process of energy
conversion of Josephson frequency to qubit excitation and photon production
in the two resonators. Note that each subunit $H_{n,m}$ can induce a
coherent quantum transition from the state $|n,m,g\rangle $ to $%
|n+1,m+1,e\rangle $; that is, each Cooper pair can tunnel to simultaneously
populate the qubit and add one photon in each cavity. Since the charge qubit
is a two-level system, its excitation will greatly inhibit the whole
system's transitions to higher occupations. Thus, the anticorrelations of
the charge transport are created: a former Cooper-pair tunneling event acts
back onto the next one; that is, a second Cooper pair can not pass through
the circuit until the flip of the qubit state. This is the key point to
induce antibunching in the photon emission.

If the system is initially prepared in the ground state $%
|0,0,g\rangle$, the Hamiltonian $%
H_{0,0}$ will dominate the dynamics, enabling a Rabi oscillation $%
|0,0,g\rangle \longleftrightarrow |1,1,e\rangle $. Obviously, the absorption of one photon in each cavity is accompanied with a excitation in the charge qubit. Since the charge qubit is treated as a two-level system, its excitation will inhibit further photon absorption, which is the mechanism involved in photon blockade \cite{PhysRevLett.118.133604,PhysRevA.97.013851}.
Consequently, the photon-pair blockade can be realized in this two-cavity
system, which offers a source of antibunched pairs of two strongly
correlated photons. It is also pointed out here that the charge qubit has finite nonlinearity in practice, and many other higher excited states exist. So, the excitation of these states is accompanied with the higher photon number excitations, which will degrade the photon blockade. However, the charge qubit is operated in the regime in which the charging energy $E_{c}$ is much larger than the Josephson coupling energy $E_{Jq}$ \cite{Bouchiat1998,Makhlin1999,Nakamura1999,Jos}. The third energy level has a eigenfrequency about $6E_{c}$, which is far greater than the transition frequency $E_{Jq}$ of the lowest two energy levels. So, the higher-order qubit excitations are greatly inhibited, which has negligible effect on the photon blockade. Compared with the previous proposal \cite{F1}, our scheme can be
implemented with the standard coplanar waveguide resonator designs, greatly
lowering the requirement for ultrahigh cavity impedance that is not
accessible in current experiment.

To check the validity of the approximation, we now investigate the dynamics
of the system by numerically solving the Schr\"{o}dinger equation with both
the full Hamiltonian $H_{I}$ and the effective Hamiltonian $H_{eff}$. With
the system initialized in the ground state $%
|0,0,g\rangle$, the time evolution of
populations $P_{00g}$ in the state $|0,0,g\rangle $ and $P_{11e} $ in the
state $|1,1,e\rangle $ is shown in Figure 2. The perfect Rabi oscillations $%
|0,0,g\rangle \longleftrightarrow |1,1,e\rangle $ are observed with both of
these two Hamiltonians $H_{I}$ and $H_{eff}$, implying that our
approximation is valid.

\subsection{Antibunched photon pair emission}

As a trigger of quantum emission of antibunched photon pairs, dissipation
has to be taken into account. When the system-environment coupling is
considered in the Born-Markov approximation, the time evolution of density
matric $\rho $ of the whole system is now governed by the master equation
\begin{equation}
\frac{d\rho }{dt}=-i[H_{eff},\rho ]+\sum_{j=1,2}\frac{\kappa }{2}%
D[a_{j}]\rho +\frac{\gamma }{2}D[\sigma _{-}]\rho ,
\end{equation}%
where $D[o]\rho =2o\rho o^{\dag }-o^{\dag }o\rho -\rho o^{\dag }o$ is the
standard Lindblad operator for a given operator $o$, and $\kappa $ ($\gamma $%
) denotes the energy damping rate of cavities (qubit). In the presence of
dissipation, a pair of two strongly correlated photons will be transferred
outside of the two cavities for each tunneling Cooper pair. We now detail
the underlying principle of the fundamental dynamics of this photon-pair
emission below.
\begin{figure}[tbh]
\centering \includegraphics[width=14.5cm]{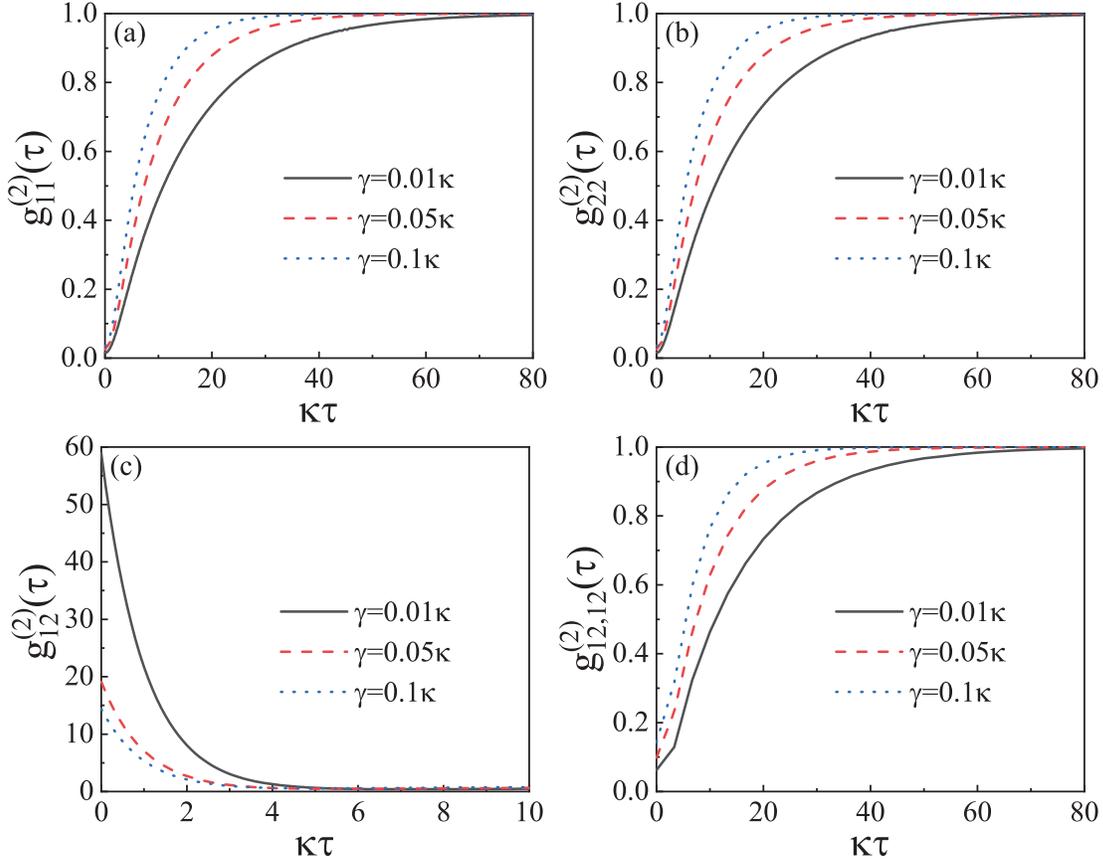}
\caption{(Color online) Steady-state photon correlation functions versus $%
\protect\kappa \protect\tau $: [a], [b] for the stand second-order
correlation functions $g_{11}^{(2)}(\protect\tau )$, $g_{22}^{(2)}(\protect%
\tau )$; [c] for the cross-correlation function $g_{12}^{(2)}(\protect\tau )$%
; [d] for the generalized second-order correlation functions $%
g_{12,12}^{(2)}(\protect\tau )$. The cavity damping rate is chosen as $%
\protect\kappa $/2$\protect\pi $ = 0.1 GHz, and the other parameters are
chosen to be the same as those in Figure 2.}
\label{fig.3}
\end{figure}

Specifically, we prepare the system initially in the ground state, and the
tunneling of a Cooper pair will draw energy quanta from the bias voltage,
inducing the coherent transition $|0,0,g\rangle \longrightarrow
|1,1,e\rangle $. Due to the energy upper limit of the two-level charge
qubit, the system populated in the state $|1,1,e\rangle $ can not be excited
to higher energy levels. This indicates that a second Cooper pair can't pass
through the circuit only after the spontaneously emission of the charge
qubit. So, to guarantee the desired antibunching, it is crucial to meet the
condition $\kappa \gg \gamma $, i.e., the coherence time of charge qubit is
much longer than that of cavities. In this situation, the two cavities will
take the lead to emit two correlated photons within the lifetime $1/\kappa $%
, stemming from the spontaneous emission of $|1,1,e\rangle $ state via the
photonic dissipation. Then, the wavefunction of the system is collapsed into
the state $|0,0,e\rangle $ but without Rabi flopping. Only after a quantum
jump $|0,0,e\rangle \longrightarrow |0,0,g\rangle $ of the qubit state
within a relative longer lifetime $1/\gamma $, a second Cooper pair can
tunnel to restart the coherent transition $|0,0,g\rangle \longrightarrow
|1,1,e\rangle $ for the next emission of a photon pair. This is the
mechanism for generating antibunched photon pairs.

On the other hand, it is worth noting here that we also can not make an
arbitrary small $\gamma$. For $\gamma\rightarrow0$, the system seems to
behave as a completely antibunched photon-pair source. However, it will take
a very long time to flip the state of charge qubit $|0,0,e\rangle
\longrightarrow |0,0,g\rangle$ and reconstruct the state $|1,1,e\rangle $
for the two cavities to emit a next correlated photon pair. This will result
in an extremely low emission rate. Therefore, there is a tradeoff between
the emission rate and the nonclassical property of the radiation field,
which can be balanced by tuning the ratio $\gamma/\kappa$.
\begin{figure}[tbh]
\centering \includegraphics[width=10cm]{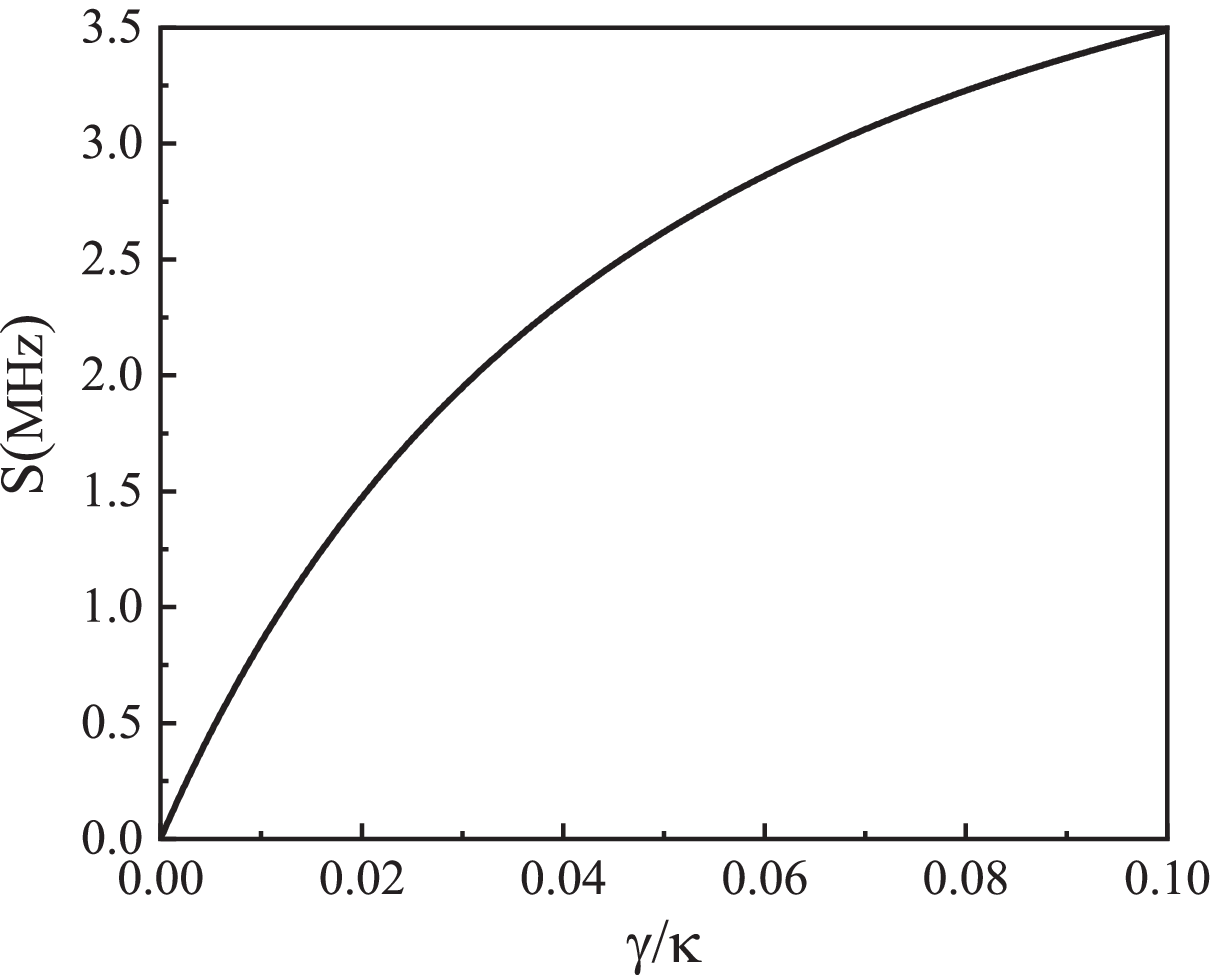}
\caption{(Color online) The photon-pair emission rate $S$ versus $\protect%
\gamma/\protect\kappa$. The relevant parameters are chosen to be the same as
those in Figure 3.}
\label{fig.4}
\end{figure}

To describe the quantum statistics of the photon emission, we further study
the following time-delay correlation functions
\begin{equation}
g_{pq}^{(2)}(\tau )=\frac{\langle a_{p}^{\dag }(0)a_{q}^{\dag }(\tau
)a_{q}(\tau )a_{p}(0)\rangle }{\langle (a_{p}^{\dag }a_{p})(0)\rangle
\langle (a_{q}^{\dag }a_{q})(\tau )\rangle }
\end{equation}%
with $p=1,2$ and $q=1,2$. For $p=q$, $g_{pq}^{(2)}(\tau )$ is just the
standard second-order correlation function that can quantify the photon
correlation emitted by a single cavity. While for $p\neq q$, it represents
the cross-correlation between the photons emitted by different cavities.
Besides, we should also introduce the generalized second-order correlation
function
\begin{equation}
g_{12,12}^{(2)}(\tau )=\frac{\langle a_{1}^{\dag }(0)a_{2}^{\dag
}(0)a_{1}^{\dag }(\tau )a_{2}^{\dag }(\tau )a_{1}(\tau )a_{2}(\tau
)a_{1}(0)a_{2}(0)\rangle }{\langle (a_{1}^{\dag }a_{2}^{\dag
}a_{1}a_{2})(0)\rangle \langle (a_{1}^{\dag }a_{2}^{\dag }a_{1}a_{2})(\tau
)\rangle },
\end{equation}%
where the joint two-photon emission event by the two cavities is considered
as a single entity \cite{Munoz2014,PhysRevA103/053710}. Here, $g_{12,12}^{(2)}(\tau )$ can capture the fundamental dynamics of photon-pair emission, and characterize the quantum
statistics of photon pairs.

In Figure 3, we plot the different steady-state correlation functions by
numerically solving the master equation (10). For $\kappa\gg\gamma$, the
zero-delay second-order correlation functions $g_{11}^{(2)}(0)\rightarrow0$
and $g_{22}^{(2)}(0)\rightarrow0$ are observed in Figures 3(a) and 3(b),
exhibiting distinct antibunching effects. So, each cavity behaves as an
excellent single-photon emitter. As seen in Figure 3(c), the zero-delay
cross-correlation function yields $g_{12}^{(2)}(0)\gg 1$. This indicates
that a pair of strongly correlated photons are emitted simultaneously by the
two cavities. Moreover, the generalized zero-delay second-order correlation
function $g_{12,12}^{(2)}(0)\ll1$ in Figure 3(d) manifests clearly that the
two cavities release their energies in the form of antibunched photon pairs.
In addition, as expected before, the changes of the different zero-delay
correlation functions indicate that our device approaches a perfect
photon-pair emitter with the decrease of $\gamma$.

Finally, we investigate the emission rate of our photon-pair source. It is
defined as%
\begin{equation}
S=\kappa \overline{n},
\end{equation}%
where $\overline{n}$ is the average photon numbers of the two cavities. The
emission rate $S$ versus $\gamma /\kappa $ is displayed in Figure 4. Under the
premise of $\kappa \gg \gamma $, we can observe that a tunable emission rate
can be achieved, i.e., $S$ gradually increases with the increase of $\gamma $%
. Hence, the emission rate can be experimentally controlled by changing the
distance between the cavity and the transmission line to adjust the value of
$\kappa $ \cite{S2}. With the currently available parameters $\omega _{1}/2\pi =9$
GHz, $\omega _{2}/2\pi =7$ GHz, $\delta /2\pi =5$ GHz, $E_{J}/2\pi =0.5$
GHz, $\lambda _{1}=\lambda _{2}=0.2$, we can obtain an emission rate of the
order of MHz.

\section{Conclusion}

In conclusion, we have proposed a practical approach to generate antibunched
photon pairs in a Josephson-photonics circuit of a dc voltage-biased
Josephson junction coupled to both a superconducting charge qubit and two
nondegenerate microwave cavities. Under an appropriate bias voltage, each Cooper pair can tunnel inelastically to cause the excitation of charge qubit and the creation of one photon in each cavity. We demonstrate that the charge
transport can be controlled via the two-level charge qubit, preventing the
simultaneous Cooper pair tunneling events. As a result, the photon-pair
blockade can be realized, i.e., the presence of a qubit excitation will impede further photon absorption. Together with the photonic dissipation, the
two cavities can emit antibunched pairs of two strongly correlated photons
with a tunable emission rate. Moreover, the generation of such a
nonclassical microwave source is compatible with current experimental
architectures, and may stimulate a variety of applications in the field of
quantum information science.

\section*{acknowledgments}
This work was supported by the National Natural Science Foundation of China
(Grant Nos. 11704306 and 12074307), the Fundamental Research Funds for the Central Universities   (Grant No. 11913291000022), and the Doctoral Scientific Research Foundation of Hubei University of Automotive Technology (Grant No. BK202113).

\appendix

\section{Detailed derivation of equation (1) in the main text}

As illustrated in Figure 1, the charge qubit is a mesoscopic superconducting island connected by a tunnel junction with capacitance $C_{q} $ and Josephson coupling energy $E_{Jq}$ \cite{Bouchiat1998,Makhlin1999,Nakamura1999,Jos}. A gate voltage $V_{g}$ is coupled to the island to control the tunneling of Cooper pairs. Its Hamiltonian is given by
\begin{equation}
  H_{q}=E_{c}(n_{q}-n_{g})^{2}-E_{Jq}\cos \phi _{q},
\end{equation}
where $E_{c}=2e^{2}/(C_{q}+C_{g})$ denotes the charging energy of a Cooper pair, and $n_{g}=C_{g}V_{g}/2e$ is the
dimensionless gate charge as a control parameter. $n_{q}$ is the number of Cooper pairs on the island, and $\phi_{q}$ is the phase difference across the junction, which are conjugate variables that obey $\left[\phi_{q}, n_{q}\right]=i$.

The $LC$ resonator consists of a capacitor $C_{j}$ connected to an inductor $L_{j}$, and its Hamiltonian is written as
\begin{equation}
  H_{r}=\sum_{j=1,2}\frac{q_{j}^{2}}{2C_{j}}+(\frac{\hbar }{2e})^{2}\frac{\phi _{j}^{2}}{2L_{j}},
\end{equation}
where $q_{j}$ and $\phi_{j}$ represent the charge and phase operators acting on the capacitor $C_{j}$ and the inductor $L_{j}$, respectively. They form a canonical conjugate pair, and satisfy the commutation relation $\left[\phi_{j}, q_{j}\right]=2ei$.

The dc-SQUID is made up of two identical Josephson junctions and can be treated as a tunable Josephson junction with effective
Josephson energy $E_{J}=2E_{J0}\cos (\pi \Phi _{\mathrm{ext}}/\Phi _{0})$, which can be tuned by controlling the magnetic flux $\Phi _{\mathrm{ext}}$ penetrating its loop area \cite{squeeze}. Here $E_{J0}$ is the Josephson junction energy of a single
junction, and $\Phi _{0}=h/2e$ is the magnetic flux quantum. The Hamiltonian of the SQUID takes the form
\begin{equation}
  H_{J}= -E_{J}\cos \phi _{J}-2en_{J}V_{J},
\end{equation}
where $\phi _{J}$ is the phase difference across the SQUID, and $n_{J}$ counts the number of Cooper pairs that
have transferred the junctions. These two variables obey the commutation relation $\left[\phi_{J}, n_{J}\right]=i$. Additionally,
$V_{J}=-\hbar\dot{\phi}_{J}/2e$ represents the voltage drop across the SQUID. In the presence of the bias voltage $V$, we have $V_{J}=V-V_{q}-V_{R1}-V_{R2}$ according to Kirchhoff's rules, where $V_{q}=-\hbar\dot{\phi}_{q}/2e$ is the voltage drop at the qubit, and $V_{Rj}=-\hbar\dot{\phi}_{j}/2e$ is the voltage drop at the $j$-th resonator. By replacing $V_{J}$ with $V-V_{q}-V_{R1}-V_{R2}$ in equation (A.3), we have
\begin{equation}
  H_{J}= -E_{J}\cos \phi _{J}-2en_{J}(V-V_{q}-V_{R1}-V_{R2}).
\end{equation}

Together with the equations (A.1), (A.2) and (A.4), we can give the total Hamiltonian of the whole system
\begin{equation}
  H_{T}=H_{q}+H_{r}+H_{J},
\end{equation}
which is just the equation (\ref{1}) in the main text.

\section*{References}
\bibliographystyle{iopart-num}
\bibliography{iopart-num}

\end{document}